\documentclass[11pt]{article}

\usepackage[T1]{fontenc}
\usepackage{graphicx}
\usepackage{amssymb}
\usepackage{color}
\usepackage{verbatim}
\usepackage{amsmath}
\usepackage{array}
\usepackage{multicol}
\usepackage{enumitem}
\usepackage{tcolorbox}
\usepackage{cancel}
\usepackage{multirow}
\usepackage{authblk}

\newcommand{\rr}{\mathbf r}

\title {\bf Electric charge redistribution in a two dimensional two component plasma for $\Gamma = 2$ induced by two impurities: a dimensional reduction }

\author[1]{Alejandro Ferrero Botero\footnote{Contact: aferrero@ucatolica.edu.co}}

\affil[1]{Departamento de Ciencias B\'asicas, Universidad Cat\'olica de Colombia\\ Bogot\'a, Colombia}

\date{}

\begin{document}

\maketitle

\begin{abstract}
In this document the density of electrically-charged positive and negative particles in a two component plasma (TCP) will be studied. Particularly, we focus on a two dimensional system confined in a large rectangular box for $\Gamma=2$ in the presence of two electric impurities. A method for solution, which will be called, {\it dimensional reduction}, will be applied in order to study the redistribution of electrically charged particles along the line joining both impurities. Numerical results, by means of a finite elements method approach, show, due to the electric field generated by the impurities, an increase in the density of charges of opposite sign in the neighborhood of each impurity. On the other hand, the presence of charges of the same sign diminishes in the same region due to the existing electric repulsion; some of the repelled particles accumulate in the border of the box. Numerical expansions around the borders of the impurities and the box show an almost linear power law relation of the net density for the particular cases that have been analyzed. It is also studied how the maximum and minimum values of the net density depend on the electric charges of the impurities, under some particular conditions. 
\end{abstract}

\vspace{2mm}


\section{Introduction}

A two component plasma (TCP) is a Coulomb system in contact with a thermal reservoir and composed by two kinds of electrical charges (ions and anions): positive charges with charge $+e$ and negative particles with charge $-e$. For instance, a salt solution with positive ions of sodium Na$^+$ and negatives of chlorine Cl$^-$ is a TCP. In two dimensions (2D), if a coupling constant (in Gaussian units) is defined as $\Gamma=\frac{e^2}{k_BT}$, where $k_B$ is the Boltzmann constant and $T$ the temperature, there exists a solvable model as $\Gamma=2$ \cite{cornu:2444}. Actually, this value defines a limit between what it is usually called weak coupling ($\Gamma<2$) and strong coupling ($\Gamma\ge 2$) \cite{Samaj1}. In the weak coupling regime, the effects induced by thermal excitations are able to avoid the collapse of oppositely-charged particles \cite{Samaj2}; however, in the strong coupling regime the plasma particles---originally treated as point particles in the continuous limit---must be considered as hard disks as such collapse cannot be avoided otherwise \cite{Kosterlitz1,Tellez1}. 

Two dimensional systems are of great interest in technology. For instance, most microscopically small devices are composed by layers of materials approximately two-dimensional  \cite{murray}. Some of such devices include semiconductors \cite{sikandar,song}, transistors \cite{Wu1}, and thin crystals \cite{horng}. Besides, the importance of graphene nowadays is evident \cite{ojeda1,ojeda2}. In physics, on the other hand, two-dimensional systems are of great interests in the study of phase transitions \cite{Zsolt}, quantum wells
\cite{Shik}, spin coupling \cite{Manchon}, among other phenomena.

In this document a two-dimensional two component plasma (2D-TCP) macroscopically neutral---same number of ions and anions in the thermodynamic limit---in the presence of two impurities will be studied for the case $\Gamma=2$. An impurity is an external particle introduced in the plasma with different properties of fugacity, charge and size than those of the ions and anions (bulk particles). Particularly, the interest of this paper is studying the charge distribution of positive and negative particles along the line joining the impurities. Some results for similar systems in the $\Gamma=2$ regime have already been studied. For example, the 2D-TCP with several adsorbing impurities modeled by a delta interaction \cite{Ferrero2007} and a 2D-TCP confined in a large disc with one electric impurity located in the center of the plasma \cite{Ferrero2014}. 

In order to study the proposed system, the model introduced by 
Cornu and Jancovici \cite{cornu:2444} will be used. Contrary to the method implemented in \cite{Ferrero2014}, in which the system respects an angular symmetry, the introduction of two impurities breaks any possible angular symmetry. This fact, in turn, translates into an impossibility of finding a coordinate system where the two variables describing the entire space can be separated---the system of differential equations describing the model cannot be solved by separation of variables. This obstacle also arises in a rectangular coordinate system, the system chosen in this document. Therefore, a method of approximation, which will be called {\it dimensional reduction}, must be introduced. 

Under this approximation method, the effects induced by one of the two directions---let it be the $y$ direction in this case---will be reduced by means of a discrete Fourier transform. Hence, the contribution arising from the $y$ coordinate is taken into account only through a sum of modes along that direction. This is analog to introducing a mean field  approximation along the Fourier-transformed coordinate. The accuracy of our results relies on assuming that the Fourier transforms of the Green functions---which are used to describe the density of ions and anions---do not have a significant dependence in the direction that has been transformed---the final solution, nonetheless, does depend on such coordinate through the sum of modes, in which the $y$-coordinate is introduced as a phase for each mode. This approximation allows to greatly reduce the complexity of the method of solution, paying the price of reducing accuracy. Additionally, this restricts the calculations of the charge distribution and correlations along horizontal lines. As previously mentioned, explicit calculations will be performed for the charge distribution along the $y=0$ line (the line that connects both impurities), which is the most relevant region. This approximation is justified by the facts that the electric fields induced by the impurities decay with the distance and, for the positions in which they are located, there is certain degree of symmetry along the two coordinate axes. In fact, the system will always be symmetric respect to the $x$ axis, while an exact symmetry about the $y$ axis only exists when the two impurities have the same electric charge.

The density and correlations of our system can be described by a set of differential equations whose solutions provide a set of Green functions---denoted as $g_{ss'}(\rr,\rr')$, where $s$ ans $s'$
refer to the two possible signs of the charges, being $\rr$ and $\rr'$ their respective positions \cite{cornu:2444}. The density of the positive particles will be denoted by $\rho_+(\mathbf r)$ and it describes the distribution of positive particles as a function of the position $\rr$ due to the electric effects of the ions and anions and an external potential (the one induced by the impurities, in this case); similarly, $\rho_-(\mathbf r)$ describes the density of negative particles. The net density will be denoted as
$\rho_T(\rr)=\rho_+(\rr)-\rho_-(\rr)$. Correlations, on the other hand, are usually denoted as $\rho^{(2)}_{ss'}(\mathbf r,\mathbf r')$ and are associated with the possibility of finding a particle of sign $s$ at $\rr$ due to the presence of a particle of sign $s'$ at $\mathbf r'$. The calculation of correlations, however, will not be of particular interest in this document.

This document is organized as follows: section \ref{sec-system} describes our system and the set of differential equations that describe it. The method of solution is described in section \ref{sec-solution}, explicitly describing the method of approximation, the set of differential equations arising from such approximation, and the numerical method implemented in subsections \ref{subsec-dim}, \ref{subsec-redy}, \ref{subsec-fin}, respectively. After showing and discussing results in section \ref{sec-results}, conclusions are stated in section \ref{sec-conclusions}.

\section{The system and formalism}\label{sec-system}

\begin{figure}
    \centering
    \includegraphics[scale=0.8]{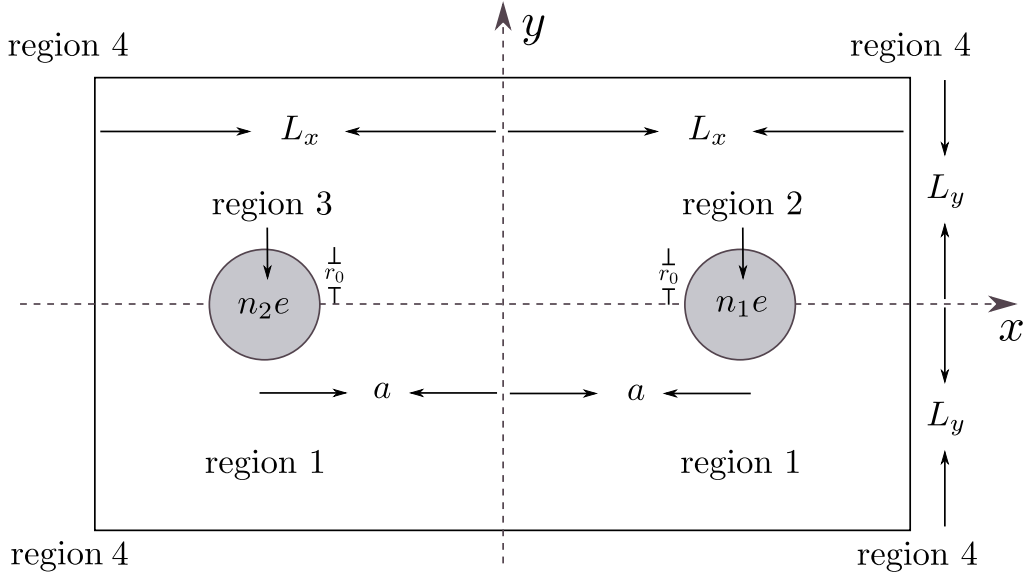}
    \caption{2D-TCP with two impurities. The line $y=0$ describes the region where the charge distribution will be computed.}
    \label{fig:plasma}
\end{figure}

Our 2D-TCP will be confined in a large rectangular box of size
$2L_x\times 2L_y$ (where $L_y\to\infty$) centered at the origin. Two impurities will be introduced, one with charge $n_1e$ located at the position $(a,0)$ and the other one with charge $n_2e$, located at the position $(-a,0)$. Although it is not strictly necessary, we will consider $n_1$ and $n_2$ to be rational numbers. A value of $n_1=3/2$ means, for instance, that three ions in the plasma are equivalent to  the charge of two impurities. The fugacity $m(\rr)$, which is associated with the kinetic contribution of the particles and models the effective pressure of the system arising form a chemical potential will be described in the following way: a) for region 1 (inside the box and outside either impurity) it will be a constant of value $m$; b) in regions 2 and 3 (inside any impurity, either the one located at the right or the left, respectively) it will be 0; and c) outside the box (region 4) it will also be 0. See figure \ref{fig:plasma} for more details.
We will use units in which $m=1$.\footnote{The fugacity $m$ represents the inverse of a correlation length, measured, as usual, in inverse-distance measure units. A value $x=4$, for instance, represents 4 correlation lengths.}

In each region $j$ ($j=1,\,2,\,3,\,4$) an electric potential will be introduced, it be denoted as $V^{(j)}(\mathbf r)$. If $n$ is the dimensionless charge of the impurity ($q=ne$), $r_0$ its radius, $\mathbf a$ a vector describing its position, and $L$ an irrelevant constant, the electric potential generated by each impurity at a position $\rr$ is\footnote{Remember that the electric potential decays logarithmically in two dimensions.}
\begin{align}\label{c:potencial}
    V(\mathbf r)= \left\{\begin{array}{lcl}-n\ln\big\vert\frac{\mathbf r-\mathbf a}{L}\big\vert &,& |\rr-\mathbf a|>r_0\,,\\
    -\frac{n}{2r_0^2}\bigl[|\rr-\mathbf a|-r_0^2\bigr]-n\ln\frac{r_0}{L}&,& |\rr-\mathbf a|\leq r_0\,.
    \end{array}\right.
\end{align}
In previous expression, the impurities are modeled as insulator discs with uniform density. For conducting discs, the potential inside the impurities is a constant whose value is $V(\rr)=-n\ln(r_0/L)$.

While an explicit form for the potential in each region will be further discussed, the main purpose in this moment is finding a set of differential equations that satisfy the Green functions in each region---for this particular system---and whose general form has been described by Cornu and Jancovici \cite{cornu:2444}. In terms of $\partial_x=\frac{\partial}{\partial x}$ and $\partial_y=\frac{\partial}{\partial y}$ and by adapting the general model to our particular system the behavior in region 1 is described by the set of differential equations
\begin{align}\label{ec:ecuac-dif1}
&\Big[\partial_x^2+\partial_y^2-m^2+2i\bigl[\partial_yV^{(1)}(\rr)\bigr]\partial_x-2i\bigl[\partial_xV^{(1)}(\rr)\bigr]\partial_y
\nonumber\\&-\bigl[\partial_xV^{(1)}(\rr)\bigr]^2-\bigl[\partial_yV^{(1)}(\rr)\bigr]^2\Big]g_{\pm\pm}^{\,(1)}(\rr,\rr')
=-m\delta(\mathbf r-\mathbf r')\,,
\end{align}
where $\delta(\mathbf r-\mathbf r')$ is the Dirac delta distribution.

On the other hand, the behavior in regions $j=2,\, 3,\, 4$ is described by the set of differential equations
\begin{align}\label{ec:ecuac-dif2}
&\Big[\partial_x \pm i\partial_y \pm \bigl[\partial_xV^{(j)}(\rr)\bigr]+i\bigl[\partial_yV^{(j)}(\rr)\bigr]\Big]g_{\pm\pm}^{\,(j)}(\rr,\rr')=0\,.
\end{align}
In terms of previous Green functions the densities for positive and negative charges can be found. They can be calculated, respectively, as
\cite{cornu:2444}
\begin{align}\label{ec:dyc}
\rho_{\pm}(\mathbf r)&=m(\mathbf r)g_{\pm\pm}(\mathbf r,\mathbf r)\,.
\end{align}

\section{Method of solution}\label{sec-solution}

\subsection{Approximation method: dimensional reduction}\label{subsec-dim}

Unfortunately, the set of differential equations described by eqs.  (\ref{ec:ecuac-dif1}) and (\ref{ec:ecuac-dif2}) cannot be solved analytically. Besides, a coordinate system that allows to perform a separation of variables ($x$ and $y$ in this case, but the same issue takes place in polar, elliptic, or any other coordinate system) cannot be implemented. This property arises as a consequence of the inclusion of the potential of the impurities. However, due to the geometry of the box, a rectangular coordinate system is the most suitable to deal with the boundary conditions. 

It is also important to recall that the system has 4 degrees of freedom: $x$, $x'$, $y$ and $y'$. The complexity inherent in solving a differential equation with 4 degrees of freedom in enormous, so suggesting the introduction of an approximation scheme. Although finite elements methods could be adapted \cite{Springer2017,DUHAMEL2007919}, the size of the linear systems expected to be obtained demands huge computational time and memory.

As previously mentioned, the problem will be faced by introducing an approximation method, named {\it dimensional reduction}. As suggested by its very name, this method attempts to reduce the number of degrees of freedom appearing in the sets of differential equations; giving up accuracy in the expected results is a fair price to pay. In this process we assume solutions of the form  $g_{ss'}^{(j)}(\rr,\rr')=\frac{1}{2\pi}\sum_{l}g_{ss'}^{(j)l}(x,x')e^{il(y-y')}$. Now, it is important to make some comments regarding the ansatz that we have chosen 
\begin{itemize}[noitemsep]
    \item $l$ is an integer number describing the individual modes. Although there are infinite modes, $l$ must be truncated from a minimum value up to a maximum one, this is done in order to obtain finite results. This maximum (minimum) value is called a cutoff, so giving rise to a regularization process. The existing divergences that demand the introduction of a regularization method, however, are inherent to the model (they will appear in any method of solution) as a consequence of the existing collapse between ions and anions previously mentioned for $\Gamma=2$ \cite{cornu:2444}.
    \item It is assumed that each individual mode $g_{ss'}^{(j)l}(x,x')$ does not depend significantly on 
    $y$ and $y'$. When a system exhibits symmetry in the $y$ coordinate, the approximation is exact (it reproduces the exact results). Although such symmetry does not exist in our problem, the fact that the electric field produced by the impurities decays with the distance suggests that such dependence is weak enough. This assumption will allow us to use the local values of the fields along horizontal lines (along regions where $y$ is constant and the Fourier-transformed modes are approximately equal). This is analog to perform a mean free field approximation along the $y$ coordinate.
    \item Performing this approximation restricts the calculation of the Green functions along horizontal lines. Actually, these are the directions in which the study of the charge distribution is most important. Due to the location of the impurities, the charge distribution along the $x$ axis will be studied. It is expected the behavior (functional form) of the densities for values such that $y>0$ to be similar; the decrease of the fields with the distance suggests that the magnitude of the effects induced by the impurities also decreases---this should not change significantly the functional form of the densities, but only the magnitude of the effects, i.e., the magnitude of the maximum and minimum values that are found.
    \item Under this approximation, the Dirac delta distribution in eq. (\ref{ec:ecuac-dif1}) can be written as $\delta(\rr-\rr')=\frac{1}{2\pi}\delta(x-x')\sum_l e^{il(y-y')}$.
\end{itemize}
A more accurate procedure does not demand the transformation of the $y$ coordinate by means of a Fourier discrete transform. On the contrary, the systems of differential equations stated in eqs. (\ref{ec:ecuac-dif1}) and (\ref{ec:ecuac-dif2}) must be solved directly, thus including the small dependence on the $y$ coordinate that has been neglected. This analysis will be developed in future studies.

\subsection{Reduction about the $y$ axis} \label{subsec-redy}

After transforming the system under the given approximation and evaluating the potentials and their derivatives---electric fields---at the given point ($y=0$), the set of differential equations that the individual modes of the Green functions, $g_{\pm\pm}^{(j)l}(x,x')$, satisfy in the different regions becomes\footnote{The complexity of the differential equations is greatly reduced because the electric fields in the $y$ direction cancel on the $x$-axis.}
\begin{subequations}
\begin{align}\label{ec:ecuac-dif1a}
&\bigg[\frac{d^2}{dx^2}
-l^2-m^2-2lE_{x}^{(1)}(x,0)
-\bigl[E_{x}^{(1)}(x,0)\bigr]^2
\bigg]g_{\pm\pm}^{(1)l}(x,x')=-m\delta(x-x')\,;
\\\label{ec:ecuac-dif1b}
&\bigg[\frac{d}{dx} \mp l \mp E_{x}^{(j)}(x,0) 
\bigg]g_{\pm\pm}^{\,(j)l}(x,x')=0\,,\,\,\,j=2,3,4\,;
\end{align}
\end{subequations}
where the fields along the line $y=0$ are given by
\begin{align}\label{ec:E1}
   E_{x}^{(1)}(x,0)&=E_{x}^{(4)}(x,0)=\frac{(n_1+n_2)x+(n_1-n_2)a}{x^2-a^2}\,,
    \\
    E_{x}^{(2)}(x,0)&=\frac{n_1}{r_0^2}(x-a)+\frac{n_2}{x+a}\,,\\
    E_{x}^{(3)}(x,0)&=\frac{n_1}{x-a}+\frac{n_2}{r_0^2}(x+a)\,.
\end{align}

\subsection{Finite elements method} \label{subsec-fin}

In order to solve the system of differential equations proposed in previous subsection, the method of finite differences \cite{Springer2017,DUHAMEL2007919} will be used. The finite elements method has been widely implemented in the scientific community. For instance, in the solution of Burgers' equations \cite{articleBurger}, the diffusion equation \cite{LIN20071533}, elliptic differential equations \cite{articleIzaian}, among others. Particularly, the simple 3-stencil approach will be implemented.

The following boundary conditions for the functions (individual modes) 
$g_{ss'}^{(j)l}(x,x')$, as defined in subsection \ref{subsec-redy}, will be used:
\begin{enumerate}[noitemsep]
    \item The individual modes vanish infinitely far from the center of the box, $g_{ss'}^{(4)\,l}(\pm\infty,x')=0$. Although we cannot implement an infinite grid numerically, the modes will vanish for values sufficiently greater than $L_x$ from the right, and values sufficiently smaller than $-L_x$ from the left.
    \item The modes are continuous at the borders of the impurities, $g_{ss'}^{(1)l}(a\pm r_0,x')=g_{ss'}^{(2)l}(a\pm r_0,x')$ and  $g_{ss'}^{(1)l}(-a\pm r_0,x')=g_{ss'}^{(3)l}(-a\pm r_0,x')$.
\end{enumerate}
By convenience, the radius of the impurity, $r_0$ and half the size of the plasma, $L_x$ will be chosen in such way that they are integer multiples of the grid size, $h$; this is done in order to be able to define the boundary conditions exactly at the required positions. 

Now, for each value of $l$ a set of differential equations must be solved using the finite elements method and using the boundary conditions just mentioned, see eqs. (\ref{ec:ecuac-dif1a}) and (\ref{ec:ecuac-dif1b}). The boundary conditions can be applied for each individual mode due to the orthogonality conditions implied by the function $e^{il(y-y')}$, which appears in the Fourier expansion of the Green function and the Dirac delta distribution.

As previously mentioned, a regularization method must be introduced in order to find finite results; otherwise, the density diverges. This is equivalent to find a maximum value for $l$ (and minimum for $-l$) for the sum of modes. This value will be denoted as $l_{max}$. As determined in \cite{cornu:2444}, this value is given by the expression  $l_{max}=e^{-\gamma}/s$, where $\gamma=0.5772\dots$ is the Euler-Mascheroni constant and $s$ is the size of the ions (and anions) of the plasma. If $s=\alpha m^{-1}$, where $\alpha$ is a positive small enough real number, we deduce that $l_{max}=\lfloor\frac{e^{-\gamma}}{\alpha}\rfloor$. We will use a value of $\alpha=0.028$, so that $l_{max}=20$.

\section{Results}\label{sec-results}

Once the functions $g_{ss'}^{(j)l}(x,x')$ are found through the method previously described, we proceed to sum up the modes along the indicated line, $y=0$. Therefore  $g_{ss'}^{(j)}(x,x')=\sum_{l=-l_{max}}^{l_{max}}g_{ss'}^{(j)l}(x,x')$. In the results shown below, the following parameters were chosen: a) the box size is $L_x=16$, b) the total grid size (this is the position where the space is truncated due to the impossibility of numerically defining and infinitely large grid) is $L_{inf}=22$, c) the distance from each impurity to the origin is $a=5.15625$ and d) the radius of each impurity is d) $r_0=1.71875$. We chose a step size of $h=0.01074$. It is useful to remember that we used units such that $m=1$.

\subsection{Calibration}

\begin{figure}
    \centering
    \includegraphics[scale=0.75]{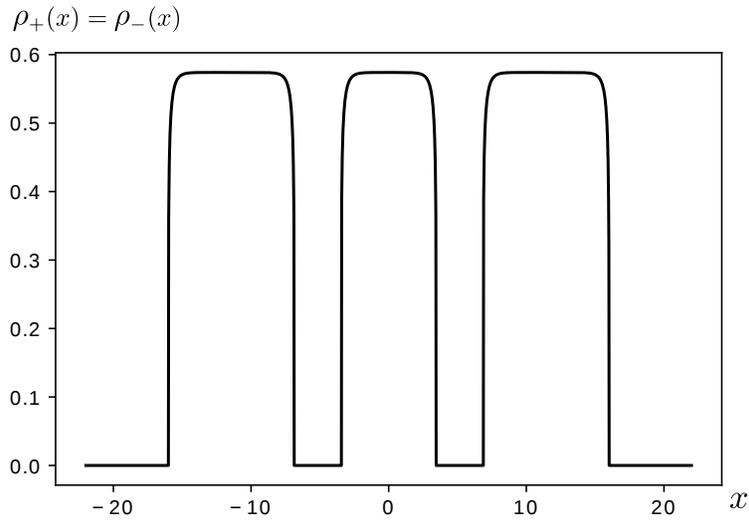}
    \caption{Charge distribution of positive and negative particles for vanishing charges---$n_1=n_2=0$---using the parameters previously stated. }
    \label{fig:nn00}
\end{figure}

Before obtaining the required results, we test (calibrate) the algorithm. In order to accomplish this, we solve the system for vanishing charges---$n_1=n_2=0$. According to \cite{cornu:2444}, the density in region 1 must be constant, while it vanishes in the other regions. This is confirmed, according to figure \ref{fig:nn00}.
In fact, for a value of $l_{max}=20$, the constant value for the density---with null charges---must be $\rho_+=\rho_-=\frac{m^2}{2\pi}(\ln(2/ms)-\gamma)=0.58752$ \cite{cornu:2444}. The obtained value was $0.57395$, so the percent error is $2.3097\%$.

\begin{figure}
    \centering
    \includegraphics[scale=0.75]{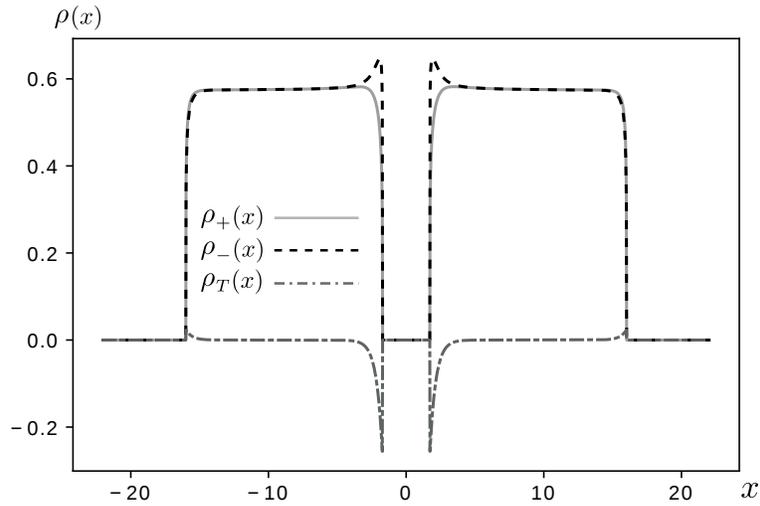}
    \caption{Distribution of positive and negative charges for an impurity with an electric charge of $n_1=2$ located at the origin. }
    \label{fig:nn10}
\end{figure}

The next step in the calibration is based on finding the charge distribution for an unique impurity, according to the analytical results found in \cite{Ferrero2014}. Nonetheless, we must perform some modifications to account for this situation, so now we use $n_1=2$, $n_2=0$ and $a=0$ (the remaining parameters remain unchanged). The results found for a unique impurity are shown in figure \ref{fig:nn10}, which agree with the expected results.
Notice how the negative charges tend to accumulate close to the impurity (with a positive charge) due to the electric attraction, while the positive charges are repelled as a consequence of the electric repulsion. Surprisingly, some amount of positive charge tends to accumulate at the borders of the box. Far enough from the borders of the box and the impurities, the density is constant and equal for both positive and negative particles (an electrostatic equilibrium is reached).

\subsection{Charge distribution for systems with impurities with non-vanishing charges}

Now we will apply the algorithm to find the net charge density along the $x$ axis as the charges of the impurities are different from zero.

It is important to point out that, for arbitrary $n_1$ and $n_2$, the density obeys the natural symmetric relation $\rho_T(x;n_1,n_2)=\rho_T(-x;n_2,n_1)$ and the antisymmetric relation $\rho_T(x;n_1,n_2)=-\rho_T(x;-n_1,-n_2)$.

\begin{figure}
    \centering
    \includegraphics[scale=0.75]{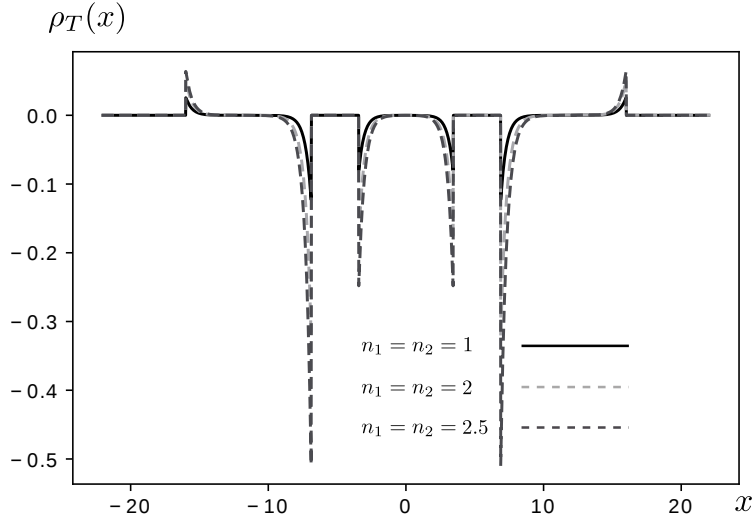}
    \caption{Net density for some impurity charges, such that $n_1=n_2$. As the signs of the charges (of the impurities) are changed, the global sign of the density also changes, due to the inherent symmetry---the shown graph is reflected about  the $x$ axis.}
    \label{fig:mama}
\end{figure}

\begin{figure}
    \centering
    \includegraphics[scale=0.75]{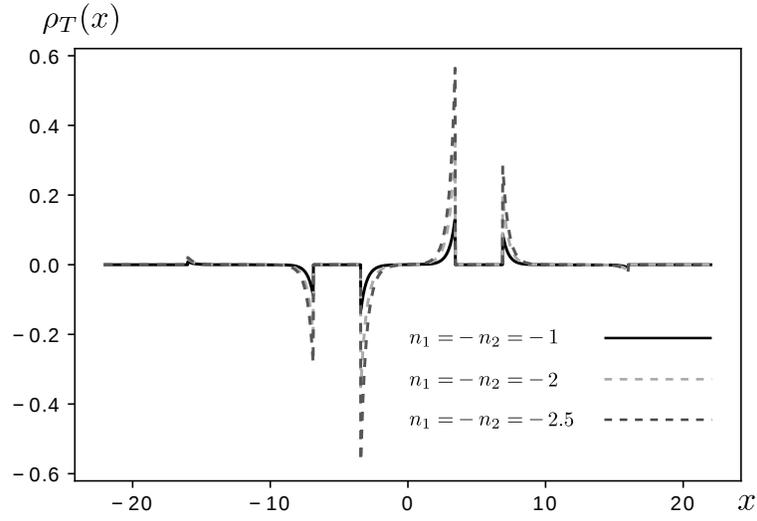}
    \caption{Net density for some impurity charges, such that $n_1=-n_2$. The same symmetry property indicated in figure \ref{fig:mama} arises as the signs of the charges of the impurities are changed.}
    \label{fig:meme}
\end{figure}

\begin{figure}
    \centering
    \includegraphics[scale=0.75]{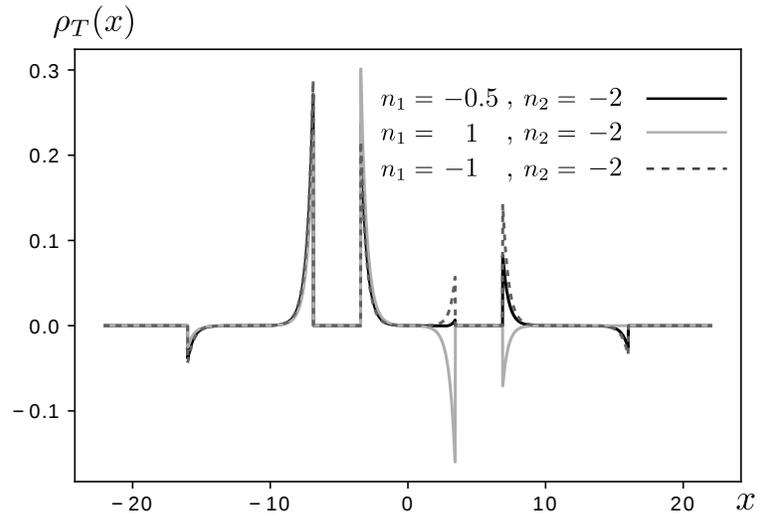}
    \caption{Net density for some impurity charges. We made $n_2=-2$.}
    \label{fig:mame}
\end{figure}

Figures \ref{fig:mama}, \ref{fig:meme} and \ref{fig:mame} show the behaviour expected. Close to the impurities charges of opposite sign tend to accumulate due to the electric attraction felt by such charges. On the other hand, equally-charged particles---than that of the impurity---are electrically repelled, so giving rise to a drastic reduction of the amount of such charges in the indicated region. It is interesting to indicate that such repelled charges spread through the plasma and some of them accumulate in the borders of the box. Last property is not unexpected; since such particles cannot leave the box, as it is not porous, the particles that are pushed to the borders have no other place to go.\footnote{Think, for instance, how a conductor in electrostatic equilibrium polarizes; positive and negative charges accumulate in the opposite sides of the conductor. Our plasma might not a perfect conductor, but particles are allowed to move under the presence of an electrostatic potential with certain degree of freedom. However, they cannot escape the box where they are confined.} The magnitude of the effects just described are stronger as the magnitude of the charges of the impurities increase.

This effect---the increase of oppositely-charged particles and the decrease of equally-charged particles---seems to obey a power law relation. In fact, according to \cite{Ferrero2014}, this behavior (for a unique impurity) is described by a superposition of modified Bessel functions. As it is well known, the modified Bessel functions of the first kind grow as $I_n(x)\sim x^n$ for short distances---$x\to0$; the ones of the second kind go like $K_n(x)\sim x^{-n}$ in the same regime---except $K_0(x)$, which decays logarithmically. For two impurities, a similar behaviour is expected when the impurities are far enough from each other, as they can almost be considered as independent particles. As $a$ is small enough, on the contrary, this is no longer valid  because the electric interaction between the impurities is significant. Particularly, for the parameters that have been studied,  approximate power law relations in the neighborhoods of the impurities and the border of the box can be obtained. Let $\rho_T^{(k)}(x)$ be such power law relation, where $k=1$ refers to the expansion around $x^{(1)}=a-r_0=3.4268$, $k=2$ to the one around $x^{(2)}=a+r_0=6.8857$, and $k=3$ to the one around $x^{(3)}=L_x=16$.\footnote{As we are interested in studying such relations for the specific cases $n_1=n_2$ and $n_1=-n_2$, expansions around $x=-a\pm r_0$ and $x=-L_x$ are related to the ones shown by symmetric and antisymmetric relations.} This relation can be written as 
\begin{align}\label{ec:expansion}
    \widetilde\rho_T^{\,(k)}(x)=\rho_T^{(k)}(x)-\rho_T^{(k)}(x_*^{(k)})=A^{(k)}|x-x_*^{(k)}|^{\alpha_k}\,,\,\,\,x^{(k)}_*\leq x\leq x^{(k)}\,,
\end{align}
where $x^{(k)}_*$ are values in the neighborhood of $x^{(k)}$, so that $|x-x^{(k)}_{*}|\leq0.214884$ and $A^{(k)}$ and $\alpha_k$ constants to be determined. Table \ref{tab:1} shows the values obtained for such constants for some values of $n_1$ and $n_2$ satisfying the relations $n_1=\pm n_2$. 
\begin{table}
\centering
{\renewcommand{\arraystretch}{1.5}
\renewcommand{\tabcolsep}{0.2cm}
\begin{tabular}{|r|r|r|r|r|r|r|r|}
\hline
$n_1$ & $n_2$ & $k$ & $x_*^{(k)}\,\,\,$ & $\rho_T^{(k)}(x_*^{(k)})\,\,$  & $A^{(k)}\,\,\,\,\,\,\,$ & $\alpha_k\,\,\,\,\,$\\
\hline
\hline
\multirow{3}{*}{1} & \multirow{3}{*}{1} & 1 & $3.2119$ & $-0.046638$  & $-0.17441$ & $1.0956$\\
\cline{3-7}
& & $2$ & $7.1006$ & $-0.073655$ &  $-0.25501$ & $1.0908$ \\
\cline{3-7}
& & $3$ & $15.780$ & $0.016160$ &  $0.048659$ & $1.0747$ \\
\hline
\hline
\multirow{3}{*}{2} & \multirow{3}{*}{2} & 1 & $3.2119$ & $-0.10637$ & $-0.38294$ & $1.0945$\\
\cline{3-7}
& & $2$ & $7.1006$ & $-0.19551$ & $-0.65229$ & $1.0921$ \\
\cline{3-7}
& & $3$ & $15.780$ & $0.032485$ & $0.097291$ & $1.0748$ \\
\hline
\hline
\multirow{3}{*}{1} & \multirow{3}{*}{$-1$} & 1 & $3.2119$ & $-0.079121$ & $-0.268681$ & $1.0897$\\
\cline{3-7}
& & $2$ & $7.1006$ & $-0.051441$ & $-0.18649$ & $1.0936$ \\
\cline{3-7}
& & $3$ & $15.780$ & $0.0052763$ & $0.015570$ & $1.0730$ \\
\hline
\hline
\multirow{3}{*}{2} & \multirow{3}{*}{$-2$} & 1 & $3.2119$ & $-0.21484$ & $-0.71149$ & $1.0920$\\
\cline{3-7}
& & $2$ & $7.1006$ & $-0.12009$ & $-0.41757$ & $1.0928$ \\
\cline{3-7}
& & $3$ & $15.780$ & $0.010564$ & $0.031157$ & $1.0731$ \\
\hline
\end{tabular}}
\caption{Parameters describing the expansions eq. (\ref{ec:expansion}) around $x^{(k)}$ for some impurity charges such that $n_1=\pm n_2$.}\label{tab:1}
\end{table}
While deducing a general relation for the expansions defined in eq. (\ref{ec:expansion}) for arbitrary values of $n_1$ and $n_2$ might not be possible (even for fixed values of $a$, $L_x$ and $r_0$) and they must probably be deduced in a case-by-case basis, it is very interesting to point out that an almost linear power law relation near $x^{(k)}$ is found in such cases. In fact, in all the cases that have been analyzed, the exponents $\alpha_k$ satisfy the relation $1.0730\leq\alpha_k\leq 1.0956$, being practically equal. Although it is clear that the larger the charges of the impurities, the more oppositely-sign particles are attracted, the slopes of the functions $\ln|\widetilde\rho_T^{\,(k)}(x)|=\ln |A^{(k)}|+\alpha_k \ln|x-x^{(k)}_*|$ around the three borders are virtually indistinguishable---only do their $y$-intercepts differ.

An important aspect that deserves a detailed analysis is the magnitude of the peaks (maxima) and/or minima that arise, both in the borders of the impurities as well as the borders of the box. In a general way, such values will be denoted as $|\rho_T^k|$, $k=1,\,2,\,3,\,4,\,5,\,6$. It is clear that the larger the magnitude of the charges of the impurities, the greater the magnitude of such peaks; this is a consequence of the electric effects---attraction and repulsion---generated by charges of the impurities, which grow proportionally with the magnitude of the charges. Nonetheless, it is worth to study such dependence in more detail. Let $|\rho_T^{1}|$, $|\rho_T^{2}|$, $|\rho_T^{3}|$, $|\rho_T^{4}|$, $|\rho_T^{5}|$ and $|\rho_T^{6}|$ denote the magnitude of such peaks---from left to right, respectively, i.e., $|\rho_T^{1}|$ represents the peak at $x=-L_x$, $|\rho_T^{2}|$ the one at $x=-a-r_0$ and so on.

Now, the notation $n_{\pm}=n_1\pm n_2$ will be used. As $n_-=0$ the charges have the same value and the system must satisfy the symmetry relation $\rho_T(x)=\rho_T(-x)$. This property is clearly observed in figure \ref{fig:mama}, where we can also deduce that  $\rho_T^1=\rho_T^6$, $\rho_T^2=\rho_T^5$ and $\rho_T^3=\rho_T^4$. On the other hand, as $n_+=0$ the charges acquire opposite signs and we now have the antisymmetry $\rho_T(x)=-\rho_T(-x)$; this relation can now be observed in figure \ref{fig:meme}, where it is also clear that $\rho_T^1=-\rho_T^6$, $\rho_T^2=-\rho_T^5$ and $\rho_T^3=-\rho_T^4$.

Based on the first graph just mentioned (the one for $n_-=0$), it is interesting to notice that  $|\rho_T^2|>|\rho^3_T|$. The reason is that, as the impurities have the same charges, the field is partially canceled in the region $|x|<a$ (there is ``destructive interference'') because the field lines generated by the impurities point in opposite directions. Conversely, in the regions such that $|x|>a+r_0$ the situation is the opposite, there is a partial addition of the field because the field lines now point in the same direction. Since the situation is opposite as $n_+=0$ (there is ``constrictive interference'' as $|x<a|$ and ``constructive interference'' as $|x|>a+r_0$), $|\rho^3_T|>|\rho^2_T|$ for this situation.

\begin{figure}
    \centering
    \includegraphics[scale=0.75]{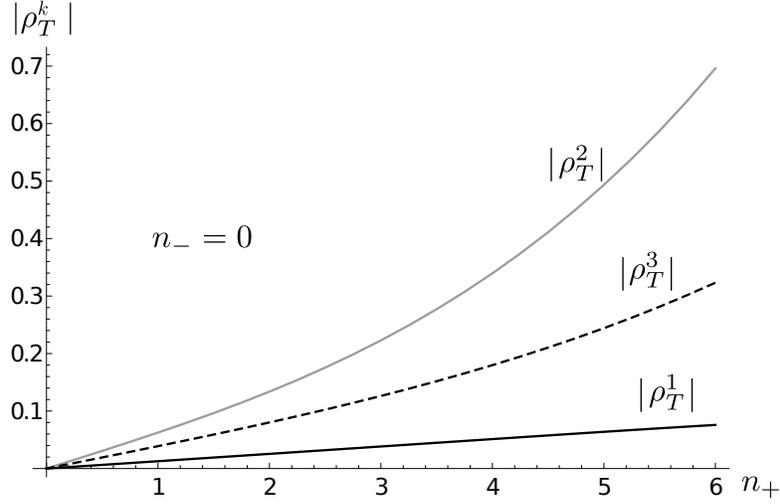}
    \caption{$|\rho_T^1|$, $|\rho_T^2|$ and $|\rho_T^3|$ as a function of $n_+$ for $n_-=0$.}
    \label{fig:picosmas}
\end{figure}

\begin{figure}
    \centering
    \includegraphics[scale=0.75]{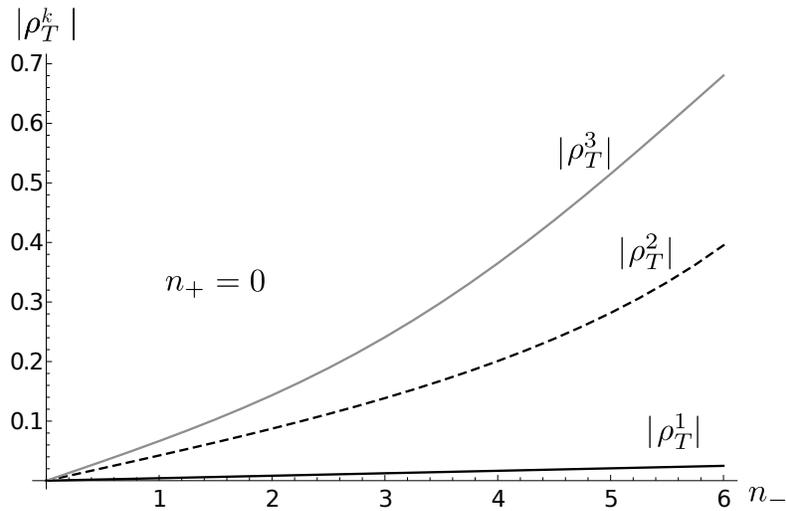}
    \caption{$|\rho_T^1|$, $|\rho_T^2|$ and $|\rho_T^3|$ as a function of $n_-$ for $n_+=0$.}
    \label{fig:picosmenos}
\end{figure}

Now, it is worth to study the magnitude of the peaks  $|\rho_T^1|$, $|\rho_T^2|$ and  $|\rho_T^3|$ for the two cases previously mentioned---$\rho_T^{k}(n_+)$ for $n_-=0$ and $\rho_T^{k}(n_-)$ for $n_+=0$. Figures \ref{fig:picosmas} and \ref{fig:picosmenos} show such behaviors.
Notice from these figures how the function $|\rho_T^{1}|$ obeys a linear relation in both cases. In fact, a numerical analysis in the interval $0\leq n_{\pm}\leq 6$ provides the following results
\begin{align}
    |\rho_T^1(n_+)|=0.012757n_+^{1.0004}\,\,\,,\,\,\,
    |\rho_T^1(n_-)|=0.0041380n_-^{1.0009}\,.
\end{align}
Nonetheless, the functions $|\rho_T^{2}|$ and $|\rho_T^{3}|$ seem to obey linear relations in the interval $0\leq n_\pm\leq 3$ (small charges), which start to increase at greater rates for larger values of $n_\pm$. Again, in the interval $0\leq n_\pm\leq 6$ we deduce the overall approximate relations
\begin{align}
    |\rho_T^2(n_+)|&=0.064891n_+^{1.2125}\,\,\,,\,\,\,
    |\rho_T^2(n_-)|=0.042974n_-^{1.1373}\,.\\
    |\rho_T^3(n_+)|&=0.039565n_+^{1.1062}\,\,\,,\,\,\,
    |\rho_T^3(n_-)|=0.069349n_-^{1.2013}\,.
\end{align}

\section{Conclusions}\label{sec-conclusions}

In this document the electric charge distribution of a 2D-TCP in the presence of two impurities has been analyzed. A method that has been called {\it dimensional reduction} along the $y$ axis and further numerical analysis by the finite elements method show the expected results, so suggesting that the such approximation is well justified. As expected, close to an impurity oppositely-charged particles accumulate due to the electric attraction, while equally-charged particles are repelled, some of them accumulating at the borders of the confining box. It was confirmed that such effects are more significant for larger  magnitudes of the charges of the impurities. The increase and decrease of the number of ions and anions close enough to an impurity follows an approximate linear power law, which, in spite that the magnitude of the peaks are clearly charge-dependent, seems to be independent of the values of the charges of the impurities---at least in the cases that have been analyzed. 

The method that has been used greatly simplifies the calculations. Although this method restricts the obtainment of the density profile to horizontal lines, the line $y=0$ is in fact the region where such results are most important to obtain. A similar behaviour is expected along other horizontal lines, taking into account that the magnitude of the peaks must be significantly reduced because of the reduction of the electric fields.

Future studies will be able to establish the behaviour of the density profile as the dimensional reduction method is not applied. Such future works will describe a more accurate behaviour of the density profile focusing on more regions within the box; besides, the amount of error associated with the approximation method that has been used can be quantified.

\section*{Acknowledgements}

This work was financed by Universidad Cat\'olica de Colombia. The author would like to thank Juan Pablo Mallarino for his suggestions, useful comments and discussions.

\bibliographystyle{ieeetr}
\bibliography{biblio}

\begin{thebibliography}{10}

\bibitem{cornu:2444}
F.~Cornu and B.~Jancovici, ``{The electrical double layer: A solvable model},''
  {\em J. Chem. Phys.}, vol.~90, no.~4, pp.~2444--2452, 1989.

\bibitem{Samaj1}
P.~Kalinay and L.~Šamaj, ``Thermodynamic properties of the two-dimensional
  coulomb gas in the low-density limit,'' {\em Journal of Statistical Physics},
  vol.~106, pp.~857--874, 03 2002.

\bibitem{Samaj2}
L.~Šamaj and I.~Travěnec, ``Thermodynamic properties of the two-dimensional
  two-component plasma,'' {\em Journal of Statistical Physics}, vol.~101, 05
  2000.

\bibitem{Kosterlitz1}
J.~M. Kosterlitz and D.~J. Thouless, ``Ordering, metastability and phase
  transitions in two-dimensional systems,'' {\em Journal of Physics C: Solid
  State Physics}, vol.~6, pp.~1181--1203, apr 1973.

\bibitem{Tellez1}
G.~Téllez, ``Equation of state in the fugacity format for the two-dimensional
  coulomb gas,'' {\em Journal of Statistical Physics}, vol.~126, 10 2006.

\bibitem{murray}
R.~Pindak and D.~Moncton, ``{Two‐dimensional systems},'' {\em Phys. Today},
  vol.~35, no.~5, p.~57, 1982.

\bibitem{sikandar}
S.~Aftab, Samiya, M.~W. Iqbal, P.~A. Shinde, A.~u. Rehman, S.~Yousuf, S.~Park,
  and S.~C. Jun, ``Two-dimensional electronic devices modulated by the
  activation of donor-like states in boron nitride,'' {\em Nanoscale}, vol.~12,
  pp.~18171--18179, 2020.

\bibitem{song}
X.~Song, J.~Hu, and H.~Zeng, ``Two-dimensional semiconductors: recent progress
  and future perspectives,'' {\em J. Mater. Chem. C}, vol.~1, pp.~2952--2969,
  2013.

\bibitem{Wu1}
P.~Wu, D.~Reis, X.~S. Hu, and J.~Appenzeller, ``{Two-dimensional transistors
  with reconfigurable polarities for secure circuits},'' {\em Nature
  Electronics}, vol.~4, pp.~45--53, 2021.

\bibitem{horng}
J.~Horng, E.~W. Martin, Y.-H. Chou, E.~Courtade, T.-c. Chang, C.-Y. Hsu, M.-H.
  Wentzel, H.~G. Ruth, T.-c. Lu, S.~T. Cundiff, F.~Wang, and H.~Deng, ``Perfect
  absorption by an atomically thin crystal,'' {\em Phys. Rev. Applied},
  vol.~14, p.~024009, Aug 2020.

\bibitem{ojeda1}
M.~Monteverde, C.~Ojeda-Aristizabal, R.~Weil, K.~Bennaceur, M.~Ferrier,
  S.~Gu\'eron, C.~Glattli, H.~Bouchiat, J.~N. Fuchs, and D.~L. Maslov,
  ``Transport and elastic scattering times as probes of the nature of impurity
  scattering in single-layer and bilayer graphene,'' {\em Phys. Rev. Lett.},
  vol.~104, p.~126801, Mar 2010.

\bibitem{ojeda2}
C.~Ojeda-Aristizabal, M.~Ferrier, S.~Gu\'eron, and H.~Bouchiat, ``Tuning the
  proximity effect in a superconductor-graphene-superconductor junction,'' {\em
  Phys. Rev. B}, vol.~79, p.~165436, Apr 2009.

\bibitem{Zsolt}
Z.~Gulácsi and M.~Gulácsi, ``Theory of phase transitions in two-dimensional
  systems,'' {\em Advances in Physics}, vol.~47, no.~1, pp.~1--89, 1998.

\bibitem{Shik}
A.~{Shik}, {\em {Quantum Wells: Physics and Electronics of Two-Dimensional
  Systems}}.
\newblock 1997.

\bibitem{Manchon}
A.~Manchon, {\em Rashba spin-orbit coupling in two-dimensional systems},
  pp.~25--64.
\newblock Netherlands: Elsevier, Jan. 2020.
\newblock KAUST Repository Item: Exported on 2020-11-11.

\bibitem{Ferrero2007}
A.~Ferrero and G.~T{\'e}llez, ``Two-dimensional two-component plasma with
  adsorbing impurities,'' {\em Journal of Statistical Physics}, vol.~129,
  pp.~759--786, Nov 2007.

\bibitem{Ferrero2014}
A.~Ferrero and G.~T{\'{e}}llez, ``Screening of an electrically charged particle
  in a two-dimensional two-component plasma at $\gamma$ = 2,'' {\em Journal of
  Statistical Mechanics: Theory and Experiment}, vol.~2014, p.~P11021, nov
  2014.

\bibitem{Springer2017}
J.~Whiteley, {\em Finite Element Methods, A Practical Guide}.
\newblock Springer, 2017.

\bibitem{DUHAMEL2007919}
D.~Duhamel, ``Finite element computation of green's functions,'' {\em
  Engineering Analysis with Boundary Elements}, vol.~31, no.~11, pp.~919 --
  930, 2007.

\bibitem{articleBurger}
T.~Ozis, E.~Aksan, and A.~Özdeş, ``A finite element approach for solution of
  burgers’ equation,'' {\em Applied Mathematics and Computation}, vol.~139,
  pp.~417--428, 07 2003.

\bibitem{LIN20071533}
Y.~Lin and C.~Xu, ``Finite difference/spectral approximations for the
  time-fractional diffusion equation,'' {\em Journal of Computational Physics},
  vol.~225, no.~2, pp.~1533 -- 1552, 2007.

\bibitem{articleIzaian}
J.~Izadian, N.~Ranjbar, and M.~Jalili, ``The generalized finite difference
  method for solving elliptic equation on irregular mesh,'' {\em World Applied
  Sciences Journal}, vol.~21, pp.~95--100, 01 2013.

\end{thebibliography}

\end{document}